\documentstyle[12pt,twoside,epsfig,fleqn,espcrc1]{article}

\title{The $\Lambda\Lambda$ interaction and the reaction 
        $\Xi^- +d \rightarrow n+\Lambda+\Lambda$
        \thanks{The authors would like to dedicate the present 
                contribution to the memory of our friend and colleague 
                Carl B.\ Dover who was instrumental in initiating this 
                investigation.}}

\author{S. B. Carr$^{\rm a}$,  
        I. R. Afnan\address{Flinders University, South Australia} and 
        B. F. Gibson\address{Los Alamos National Laboratory, NM, U.S.A.}}

\begin{document}

\maketitle

Interest in the $\Lambda\Lambda$ interaction is partly due to the
presence of quark-model predictions for an $S=-2$ dibaryon \cite{J77}, and
partly to the interest in the role of the coupling between the
$\Lambda\Lambda$ and $\Xi N$ channels in $\Lambda\Lambda$
hypernuclei \cite{G94}. This latter effect is expected to be substantially 
more important than the coupling of the $NN$ to the $N\Delta$ channel in
the  $S=0$ sector, since the difference in threshold between the
$\Lambda\Lambda$ and $\Xi N$ is only $\approx$25~MeV. In the absence
of any direct measurement of the $\Lambda\Lambda$ amplitude, we must 
resort to either $\Lambda\Lambda$ hypernuclei, or to a reaction with
a $\Lambda\Lambda$ final-state interaction to determine the $YY$
(the $S=-2$ baryon-baryon system) interaction. 
In this report we present results of a theoretical 
study of the hypernucleus $^{\ \; 6}_{\Lambda\Lambda}$He and the 
reaction $\Xi d\rightarrow n\Lambda\Lambda$ whereby 
we examine the sensitivity of the calculations to details of the 
$\Lambda\Lambda$ potential, and the coupling between the
$\Lambda\Lambda$ and the $\Xi N$ channels.

\begin{figure}[th]
\begin{minipage}[b]{.49\linewidth}
  \centering\epsfig{figure=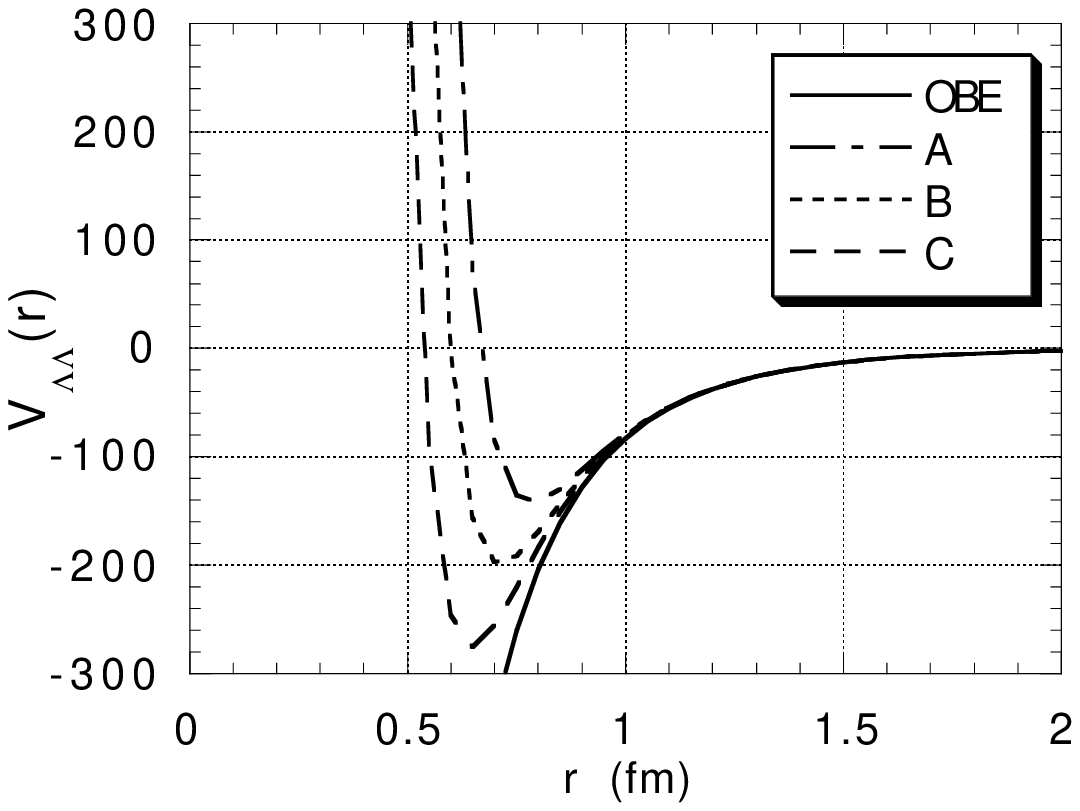,width=\linewidth}
\end{minipage}\hfill
\begin{minipage}[b]{.49\linewidth}
  \centering\epsfig{figure=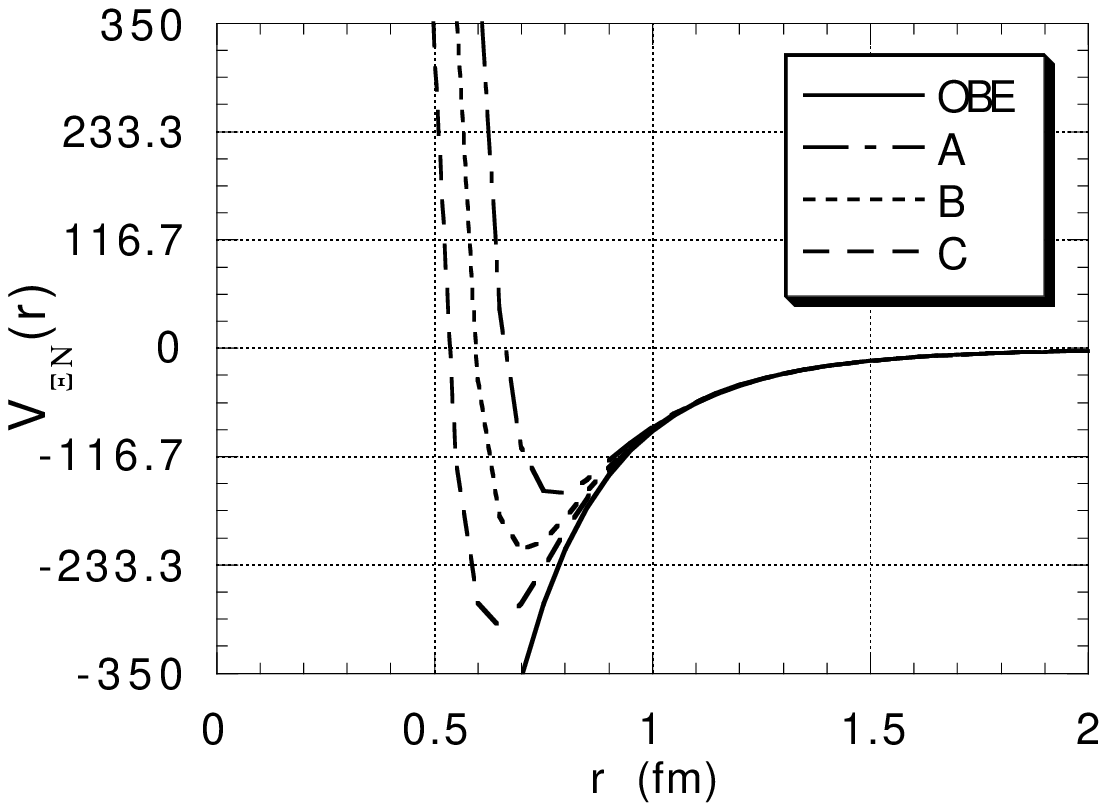,width=\linewidth}
\end{minipage}
\caption{The $^1$S$_0$ OBE $\Lambda\Lambda$ and $\Xi N$ potentials 
with and without short range repulsion.}\label{fig.1}
\end{figure}

In the absence of any data on the $YY$ interaction, we have taken the
meson exchange part of the Nijmegen $D$ potential for the $NN$ 
system \cite{NRS77}, and performed an $SU(3)$ rotation to determine the 
couplings of the mesons to the hyperons. For a purely $S$-wave 
interaction, the one boson exchange potential (OBEP) for the exchange
of the $i^{\rm th}$ meson is given by
\begin{equation}
V_i(r) = V^{(i)}_c(r) 
   + \vec{\sigma}_1\cdot\vec{\sigma}_2\ V^{(i)}_\sigma(r)\ .\label{eq:1}
\end{equation}
Since the resulting OBEP is singular at the origin, we introduced a repulsive 
soft core with a cut-off mass $M\approx 2.5$~GeV. As a result the radial 
potential for the exchange of the $i^{\rm th}$ meson is:
\begin{equation}
V_{\alpha}^{(i)}(r) = V_0^{(i)} \left[\frac{e^{-m_ir}}{m_ir} 
          - C\,\left(\frac{M}{m_i}\right)\, 
                     \frac{e^{-M r}}{M r}\right]
            \qquad \alpha = c,\sigma             \ ,\label{eq:2}
\end{equation}
where $m_{i}$ is the mass of the exchanged meson and $V_{0}^{(i)}$ 
is given in terms of masses and  coupling constants as determined 
by the $NN$ data \cite{C96,CAG97}. The parameters $M$ and $C$ are adjusted to 
ensure that the long range part $(r>0.8\ {\rm fm})$ of the meson 
exchange potential is not modified by the choice of cut-off. The final 
$\Lambda\Lambda$--$\Xi N$ interaction in the $^1$S$_0$ channel has been
chosen to either support a bound state (C), generate an anti-bound state (B), 
or have no bound state at all (A). In this way we can test the hypothesis 
that the $^1$S$_0$ $\Lambda$--$\Lambda$ interaction is comparable in 
strength to the $^1$S$_0$ $n$--$n$ potential \cite{Dover}. In 
Fig.~\ref{fig.1} we illustrate the diagonal elements of 
the coupled channel potentials and include the OBEP with no cut-off for 
comparison.

\begin{table}[thb]
\caption{The effective range parameters for the local and separable
potentials in the $^1$S$_0$  $\Lambda\Lambda$--$\Xi N$.}
\begin{tabular}{lccccc} \hline
 Pot. & $a_{\Lambda\Lambda}$~(fm) & $r_{\Lambda\Lambda}$~(fm) & 
$a_{\Xi N}$~(fm) & $r_{\Xi N}$~(fm) & B.E. (MeV) \\ \hline
    A    & -1.91 & 3.36 & -2.12-0.75i & 3.45-0.45i & UB   \\
   SA    & -1.90 & 3.33 & -2.08-0.81i & 3.44-0.22i & UB   \\ \hline
    B    & -21.1 & 1.86 & -2.05-6.53i & 2.12-0.21i & UB   \\
   SB    & -21.0 & 2.54 & -2.07-6.52i & 2.62-0.15i & UB   \\ \hline
   C1    & 7.82  & 1.41 & 3.08-5.26i  & 1.74-0.144i& 0.71 \\ 
  SC1    & 7.84  & 1.48 & 3.05-5.28I  & 1.45+0.074I& 0.71 \\ \hline
   C2    & 3.37  & 1.0  & 3.37-2.54i  & 1.44-0.10i & 4.74 \\
  SC2    & 3.36  & 1.0  & 3.35-2.50i  & 1.83-0.09i & 4.73 \\ \hline 
\end{tabular}
\end{table}

Since the above procedure gives a local coordinate space potential
and because we propose to carry through a three-body calculation for the 
hypernucleus $^{\ \; 6}_{\Lambda\Lambda}$He and the breakup reaction
$\Xi^- d\rightarrow n\Lambda\Lambda$, we have constructed a set of 
$S$-wave separable potentials that give the same scattering length 
and effective range as the local potentials in the $S=-2$ sector. In Table~1
we present the effective range parameters in the $^1$S$_0$ 
$\Lambda\Lambda$--$\Xi N$ channel for the local and separable 
potentials. Here we note that potential SA has no bound state, 
potential SB has a virtual or anti-bound state, while the potentials SC1 
and SC2 give a binding energy of 0.71 and 4.74~MeV.
To test the resulting potentials with the only piece of
experimental data on the $\Lambda\Lambda$ interaction, we have
calculated the binding energy of $_{\Lambda\Lambda}^{\ \; 6}$He as
a $\alpha\Lambda\Lambda$ three-body system using the Alt-Grassberger 
Sandhas equations \cite{AGS67}. To examine the role of 
the coupling in the $\Lambda\Lambda$--$\Xi N$ channels, we have 
performed three distinct calculations by: 
(i)~Including the coupling between the channels and solving the 
equation for the $\alpha\Lambda\Lambda$--$\alpha\Xi N$ system. 
(ii)~Discarding the coupling between the channels without any
modification to the parameters of the potential. 
(iii)~Excluding the coupling between the channels, but adjusting 
the parameters to give the same $\Lambda\Lambda$ effective range 
parameters as the corresponding local potential. 
The results are presented in Table~2.
\begin{table}[th]
\caption{The binding energy in MeV of $^{\ \; 6}_{\Lambda\Lambda}$He for the four
  potentials under consideration.}
\begin{tabular}{l|cccc|c} \hline
              & SA    & SB    & SC1    & SC2    & Exp. \\ \hline
$\alpha\Lambda\Lambda$ -- $\alpha\Xi N$ & 
9.738 & 12.268 & 15.912 & 19.836 & \\
$\alpha\Lambda\Lambda$ with no coupling to $\alpha\Xi N$ &
9.508 & 11.606 & 14.533 & 17.508 & 10.9$\pm$ 0.6 \\
$\alpha\Lambda\Lambda$ with effective $\Lambda\Lambda$ potential & 
10.007 & 14.134 & 17.842 & 23.750 & \\ \hline 
\end{tabular}
\end{table}

If we compare these binding energies for $_{\Lambda\Lambda}^{\ \; 6}$He
with the one experimental measurement of $10.9\pm0.6$~MeV, we find 
that: 
(i)~By comparing row one and three of Table~2, we may conclude
that the inclusion of the coupling at the two-body level is essential if we
are to avoid over-binding in heavier nuclei.
(ii)~From row one and two we observe that the contribution of the 
coupling between the $\Lambda\Lambda$ and $\Xi N$ in 
$^{\ \; 6}_{\Lambda\Lambda}$He is small. This is due to the fact that 
the nucleon in the $\alpha\Xi N$ Hilbert space is Pauli blocked. 
(iii)~The results in Table~2 suggest that the potential SB predicts
the result closest to the experimental separation energy, and
therefore best represents the $\Lambda\Lambda$ interaction. 
This supports the suggestion that the $\Lambda$--$\Lambda$ $^1$S$_0$ 
interaction strength may in fact be comparable to that of the 
$n$--$n$ $^1$S$_0$ interaction.
\begin{figure}[hbt]
\begin{minipage}[b]{.49\linewidth}
  \centering\epsfig{figure=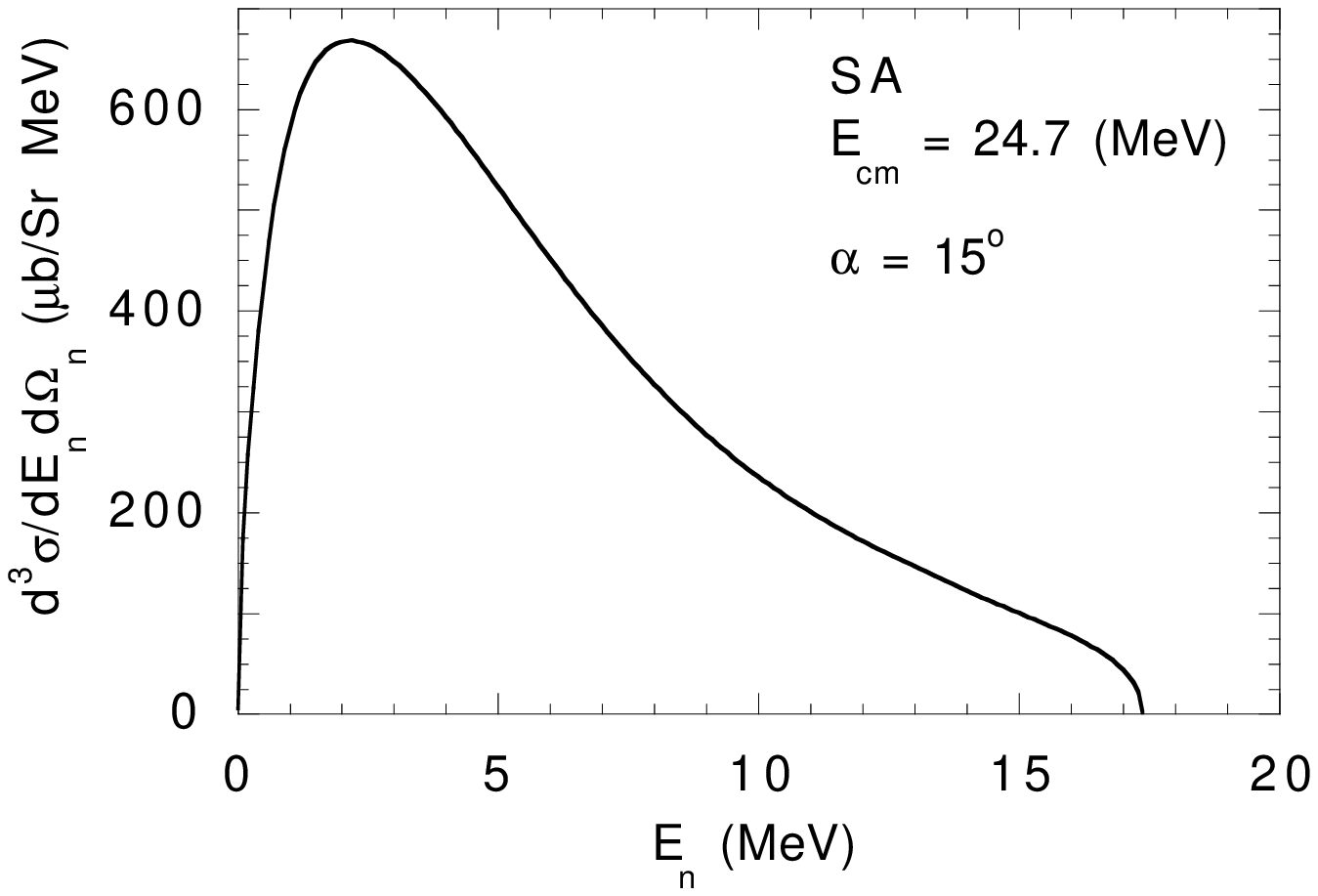,width=\linewidth}
\end{minipage}\hfill
\begin{minipage}[b]{.49\linewidth}
  \centering\epsfig{figure=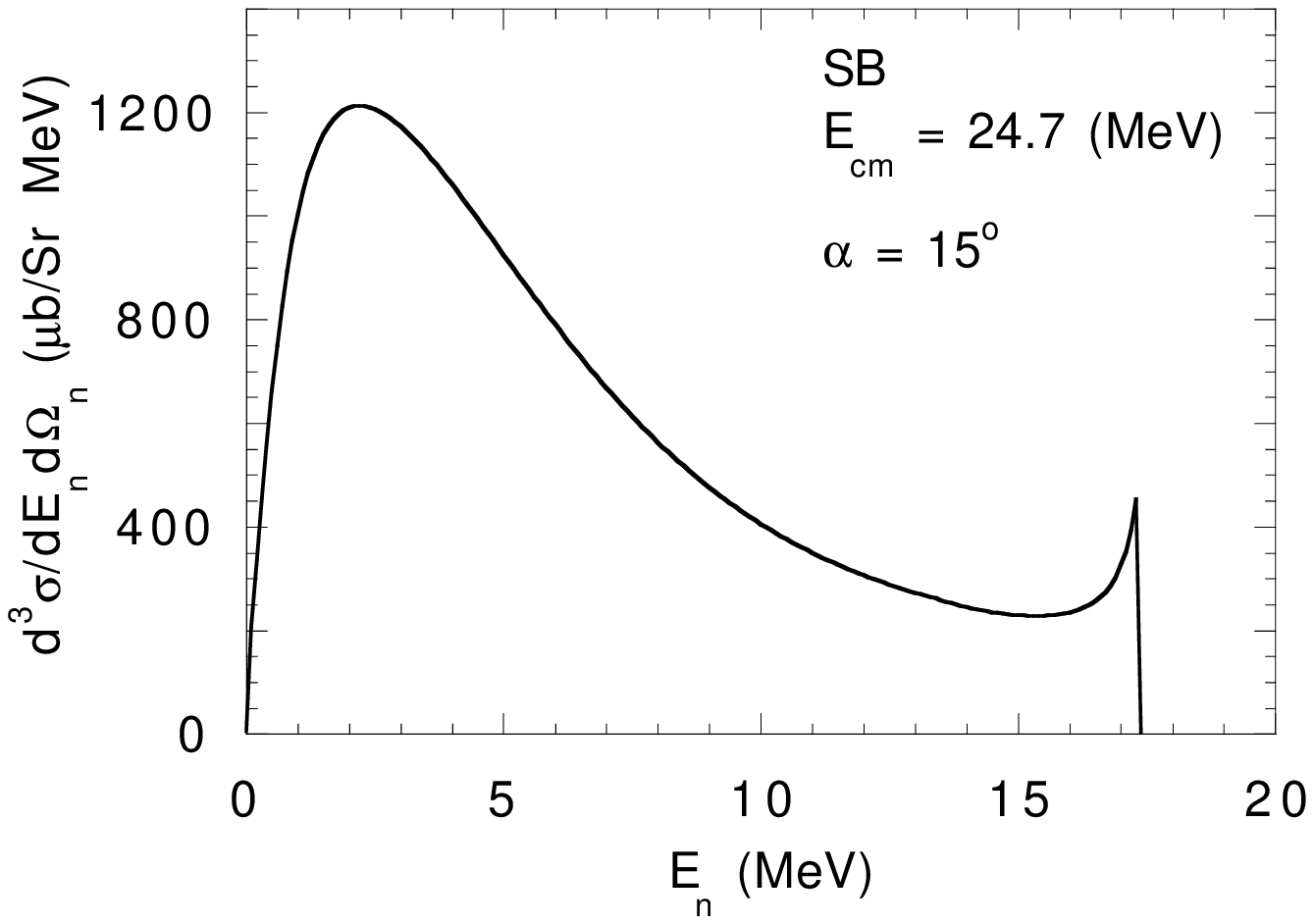,width=\linewidth}
\end{minipage}
\begin{minipage}[b]{.49\linewidth}
  \centering\epsfig{figure=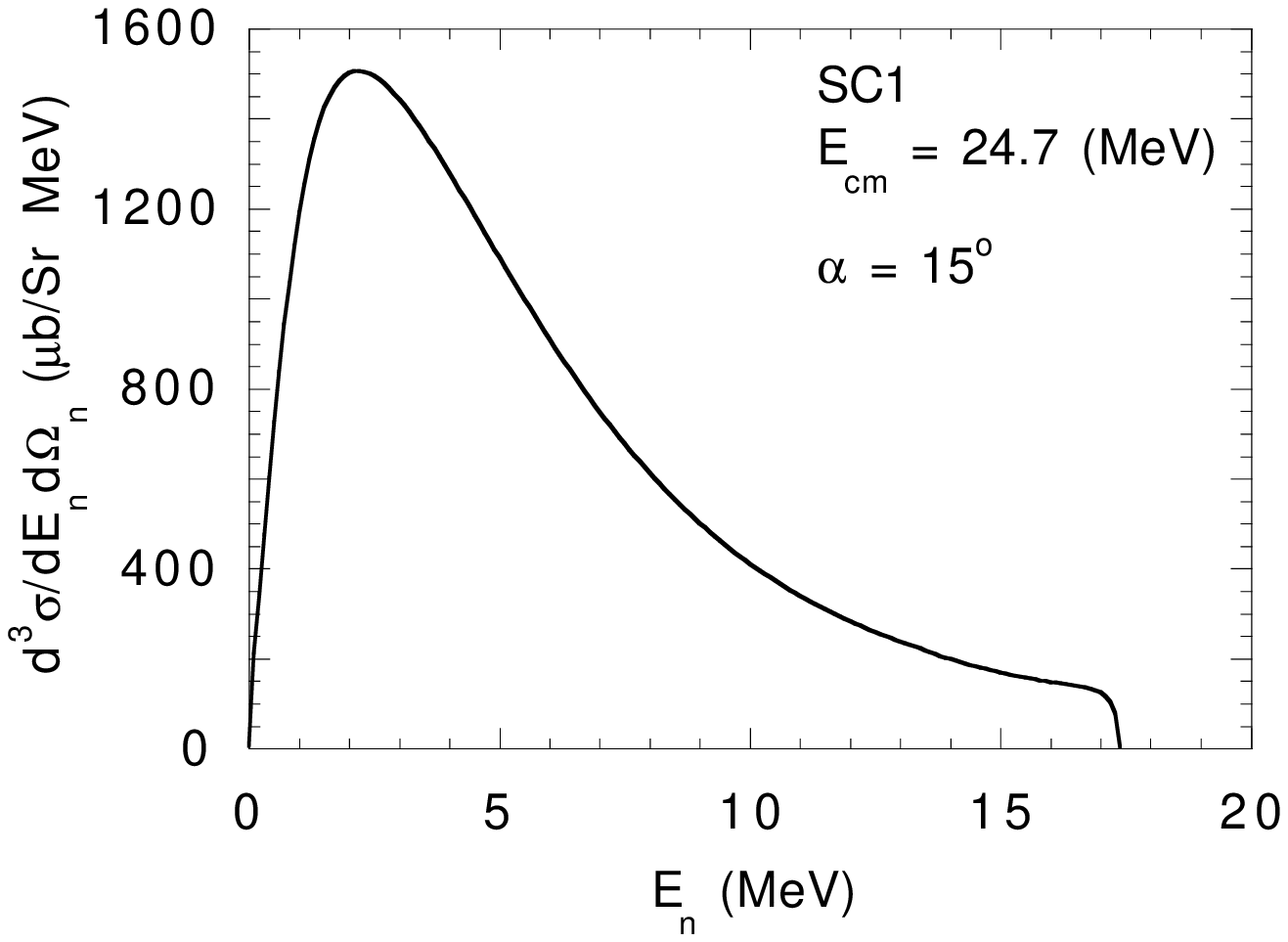,width=\linewidth}
\end{minipage}\hfill
\begin{minipage}[b]{.49\linewidth}
  \centering\epsfig{figure=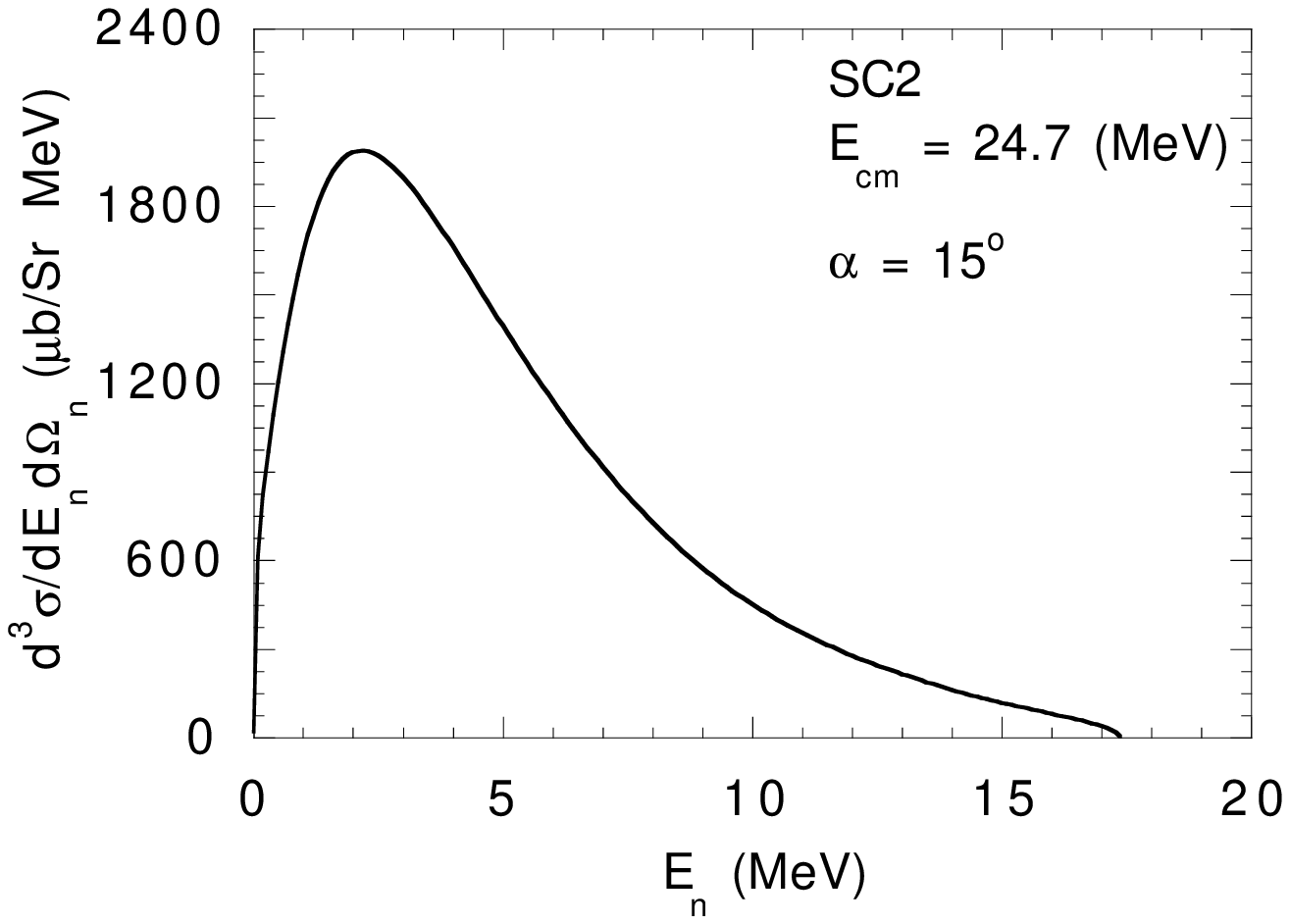,width=\linewidth}
\end{minipage}
\caption{The NDES for the potentials SA, SB, SC1, and 
SC2.}\label{fig.2}
\end{figure}

We now turn to the reaction $\Xi^-d\rightarrow n\Lambda\Lambda$ for
which there is an experiment in progress at Brookhaven \cite{B92}. In the 
Fig.~\ref{fig.2}  we show the neutron differential energy spectrum (NDES) 
for this reaction for the four potentials under consideration. The 
energy at which the calculations have been performed corresponds to 
the $\Xi^-$ capture by the deuteron at rest.  With the
exception of the result for the potential SB, the neutron 
spectrum does not exhibit the final state interaction (FSI) peak expected. 
In all four cross sections the dominant feature is the large broad
peak at the low-energy end of the neutron spectrum. 

\begin{figure}[hbt]
\centering\epsfig{figure=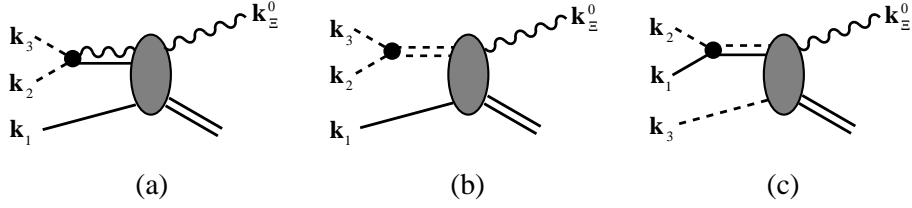,width=12cm}
\caption{The three amplitudes that contribute to the NDES for 
$\Xi^- d\rightarrow n\Lambda\Lambda$.}\label{fig.3}
\end{figure}

\begin{figure}[ht]
\begin{minipage}[b]{.49\linewidth}
  \centering\epsfig{figure=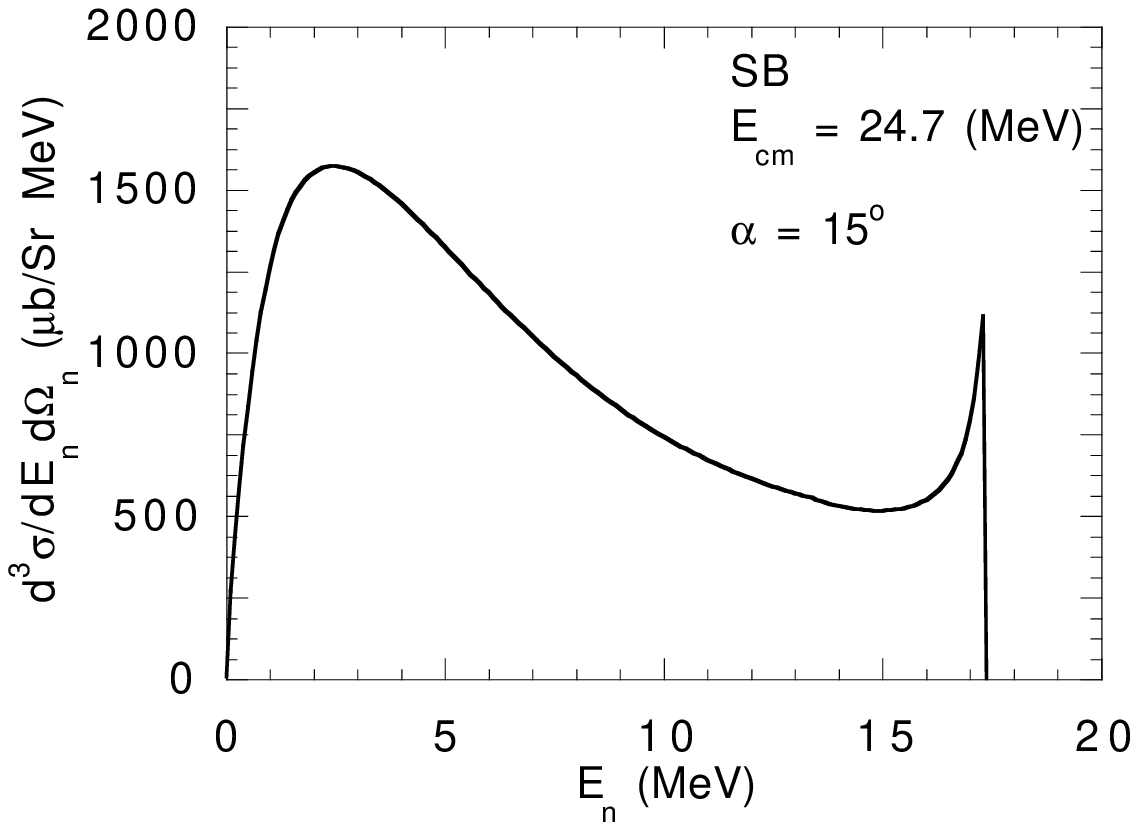,width=\linewidth}
  \caption{The NDES for (a) plus (c). }\label{fig.4}
\end{minipage}\hfill
\begin{minipage}[b]{.49\linewidth}
  \centering\epsfig{figure=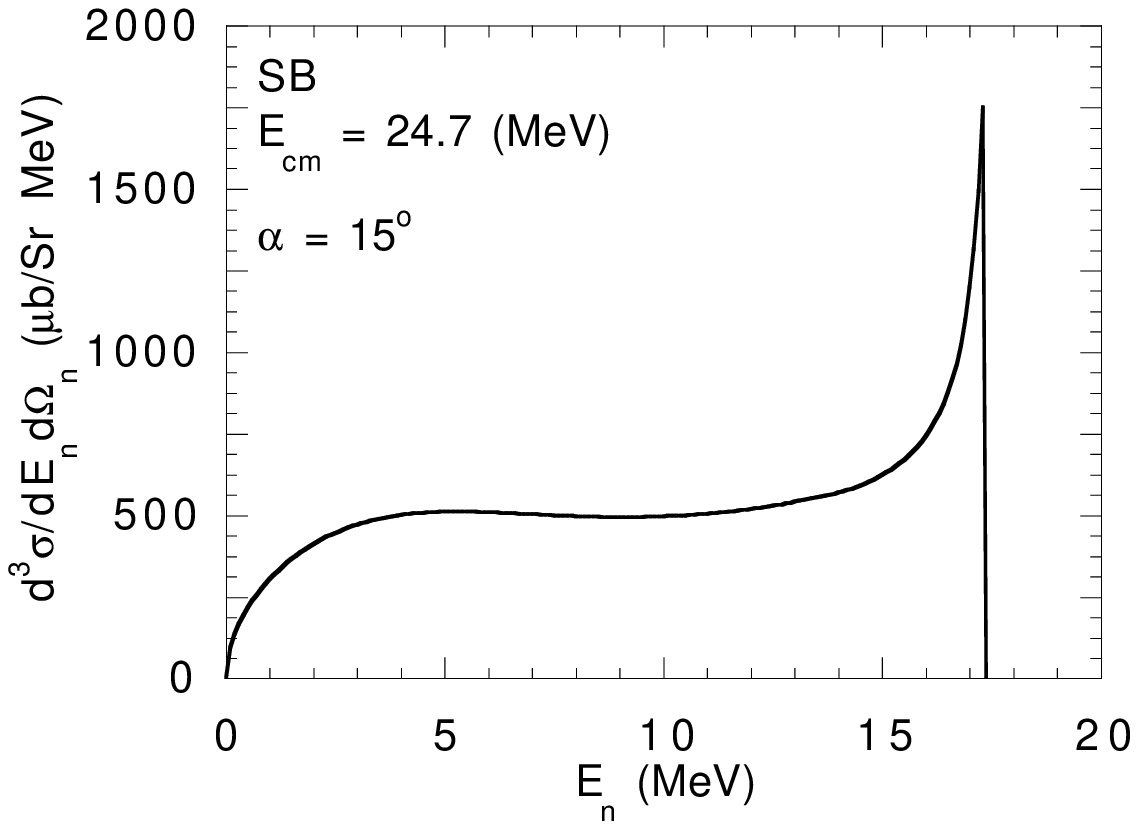,width=\linewidth}
  \caption{The NDES for (b) plus (c).}\label{fig.5}
\end{minipage}
\end{figure}

A detailed investigation of the different contributions to the NDES 
reveals that the suppression of the FSI is the result of a destructive 
interference between the amplitudes that contribute to the NDES.
In Fig.~\ref{fig.3} we give a diagrammatic representation of the three
amplitudes that contribute to the cross section. Diagrams (a) and (b)
are expected to contribute to the FSI peak, since the
final interaction is in the $\Lambda\Lambda$--$\Xi N$ coupled channels
which for the potential SB is dominated by the anti-bound state
pole. On the other hand, diagram (c) is a background term that could
interfere constructively with either of the amplitudes corresponding 
to diagrams (a) and (b).
To determine the relative sign of the three amplitudes we present in
Figs.~\ref{fig.4} and \ref{fig.5} the NDES for the diagrams (a) plus (c) 
and (b) plus (c) respectively. Here from the height of the FSI peak, we 
may conclude that diagrams (a) and (c) interfere destructively, while 
diagrams (b) and (c) give an enhancement in the FSI peak. This implies 
that diagrams (a) and (b) are out of phase. Since both of these diagrams 
are dominated by the anti-bound state in the FSI region, the fact 
that they are out of phase implies that in  the cross section the 
FSI peak is suppressed. This suppression of the FSI peak is purely 
the result of the fact that the final interaction is in the 
$\Lambda\Lambda$--$\Xi N$ coupled channel for which the diagonal
$\Lambda\Lambda\leftarrow\Lambda\Lambda$ $T$-matrix is out of phase
with the non-diagonal $\Lambda\Lambda\leftarrow\Xi N$ $T$-matrix and
as a result we have a cancellation between diagrams (a) and (b). 
This is to be compared with the $n-d$ break-up where the final $n$--$n$ 
interaction is a single channel $^1$S$_0$ with a resultant enhancement
in the FSI peak.

From Figs.s~\ref{fig.4} and \ref{fig.5} we may also deduce that the broad 
peak at the low-energy end of the neutron spectrum comes from diagram (a). 
In lowest order, this is 
proportional to the momentum distribution of the neutron in the
deuteron, {\it i.e.} the momentum space deuteron wave function. Since 
we used the same wave function with the four potentials under 
consideration, the shape of this peak in the NDES is the same for the
four potential (see Fig.~\ref{fig.2}).

From the above results for the NDES we may conclude that in the event
of an $S=-2$ dibaryon being present just below the $\Lambda\Lambda$
threshold, it would give rise to a clean signal not to be confused with a 
FSI peak.

\end{document}